# $HfO_{2-x}$ Neuromorphic Memristor Based on Tantalum and Molybdenum Electrodes

Kerem Karatas[1*], Bünyamin Özkal[1,3], Abdullah H. Cosar[2], Sinan Kazan[1]

*[1]Department of Physics, Gebze Technical University, Kocaeli, Turkiye*

*[2]Department of Materials Science, Gebze Technical University, Kocaeli, Turkiye*

*[3]Advanced Materıals and Research Laboratory, MKE, Ankara, Turkiye*

## ABSTRACT

In this research, we report the fabrication and characterization of a Ta/$HfO_{2-x}$/Mo memristor. The synaptic behavior of a Ta/$HfO_{2-x}$/Mo memristor has been investigated. $HfO_{2-x}$ (15 nm) was grown using the pulsed laser deposition (PLD) method. Electrodes were fabricated using a sputtering system and photolithography method. The metal oxide stoichiometry was ascertained via X-ray photoelectron spectroscopy (XPS). A pinched hysteresis loop is demonstrated through I-V measurements, confirming SET and RESET states. I–V measurement is supported by conduction mechanism analysis. A clear separation between high- and low-resistance states is shown during repeated reads. The neuromorphic characteristics of the device are confirmed through long-term potentiation/depression and paired-pulse facilitation. These results demonstrate that Ta/$HfO_{2-x}$/Mo is a promising configuration for resistive switching device research and neuromorphic applications.

**KEYWORDS:** Memristor, Synaptic Memristor, Resistive Switching, $HfO_2$, Tantalum, Molybdenum, PLD

***Corresponding author:** keremkaratas@gtu.edu.tr

## INTRODUCTION

Artificial intelligence (AI) aims to create systems capable of human-like thinking and behavior, encompassing abilities such as perception, reasoning, learning, planning, and prediction [1,49]. Significant advantages are achieved in different fields such as automation & robots, the automotive industry, finance, healthcare, and daily applications [2]. These performances demand high energy consumption, rapid processing, and massive data storage [3]. When examining von Neumann architecture, the foundational model of current computers must satisfy the requirements for low power consumption, high speed, and excessive data storage [4-8]. Since the von Neumann architecture separates the memory functions and computing processes, this causes complexity and a substantial computational workload on the computers' Central Processing Unit (CPU) [9]. Therefore, it is essential to shift toward neuromorphic computing. Neuromorphic computing is a method that aims to develop computer systems that can mimic the structure and functioning of the human brain. Emulating the human brain enables systems to replicate key brain functions such as learning, processing, analysis, and memory formation [10-12]. Artificial neural networks (ANNs) play a central role in neuromorphic computing. These are models that simulate the brain's structure and function through layers of interconnected units called neurons. Biological neurons transmit electrical signals and connect with other neurons through synapses, forming the basis of learning and memory processes. Neuromorphic computing uses electrical devices to mimic how the human brain works by replicating the interactions between neurons and synapses [13-14].

Memristors, proposed by Leon Chua in 1971 [15], are resistive switching devices capable of emulating synaptic functions, such as learning, memory retention, and adaptability. They possess essential characteristics for neuromorphic systems, including low power consumption, high speed, non-volatile memory, and multi-level data storage. These devices change their resistance states in response to electrical stimuli, allowing them to store and process information similarly to biological synapses.

This dynamic resistance adjustment enables memristors to replicate crucial processes of learning and adaptation, making them valuable for neuromorphic computing [16-28]. Several studies have alreadty demonstrated that the memristors can be used as both artificial synapses and artificial neurons, enabling neuromorphic systems to be designed at the hardware level more efficiently and closer to biological neural networks [57-59].

In the processes of fabrication and use of memristors and the development of memristor technology, materials science plays a significant role [60]. Metal oxides such as hafnium oxide ($HfO_2$) and titanium oxide ($TiO_2$) enable memory functionality by providing hysteresis-based resistance switching and are preferred for their high switching stability, wide bandgap, and oxygen vacancies that allow controlled resistance changes [47,61-63]. Materials like tantalum (Ta), molybdenum (Mo), silver (Ag), nickel (Ni), and platinum (Pt) are preferred due to their conductivity, chemical stability, oxidation resistance, and material compatibility for electrode integration [40,54-56].

$HfO_{2-x}$ is a material worth investigating for memristor production due to its distinctive characteristics. $HfO_{2-x}$ based memristors are characterized by their scalability, stability (under different thermal and environmental conditions), controllability, and efficient resistive switching properties [29-31]. Its oxygen-deficient structure enables controlled formation of oxygen vacancies to realize memristive behaviors. Its analog switching characteristics simulate the analog nature of synapses, where the changes in resistance states imitate the synaptic weights. The configuration of $HfO_{2-x}$ with metallic layers, such as Ta and Mo leads to enhanced switching while modifying the effective band structure [32-38] due to their strong oxidizing tendency [44]. Therefore, Ta/$HfO_{2-x}$/Mo is a structure worth exploring for revolutionary steps in neuromorphic computing.

In this work, we have fabricated and investigated the Ta/$HfO_{2-x}$/Mo-based memristor device on a $SiO_2$ wafer. Ta and Mo electrodes were used as bottom and top electrodes fabricated by RF/DC magnetron sputter using standard photolithography technique for $HfO_{2-x}$ based synaptic-memristor device. $HfO_{2-x}$ layer was grown using the Pulsed Laser Deposition (PLD) technique [47]. In this technique, high-k oxide materials are used as a single crystal target, allowing the growth of crystalline film. The laser can vaporize nanoparticles from the target in a thin film on the substrate than other growth methods at optimum oxygen [45] and substrate temperature [46] values. Using X-ray photoelectron spectroscopy (XPS), the stoichiometry of the $HfO_{2-x}$ was examined. Memristive behavior has been performed by the electrical characterization of Ta/$HfO_{2-x}$/Mo memristor, which showed a nonlinear I-V curve exhibiting a pinched hysteresis loop. The I–V measurements were supported by set-state fitting of the device using Fowler–Nordheim tunneling, space-charge-limited conduction (SCLC), and Schottky emission models. The device's capacity to transition between a high resistance state (HRS) and a low resistance state (LRS) is necessary for both artificial synapse and memory applications. The retention tests further highlighted the device's potential for long-term data storage by showing it could sustain steady LRS and HRS values over time.

These devices can also be used in neuromorphic systems, where they can mimic the plasticity of biological synapses and allow AI systems to perform advanced cognitive functions, thanks to the observation of long-term depression (LTD) and long-term potentiation (LTP) synaptic behaviors in these systems [64].

## EXPERIMENTAL DETAILS

Photolithography, DC sputter, and PLD thin-film manufacturing have been used in the device's fabrication. The SUSS MBJ4 mask aligner has been used with AZ 1505 positive photoresist and a chrome mask to obtain the Mo top electrode (TE) lithographically. BE with a 15 nm Ta layer was produced on a thermally oxidized Si substrate in a high-vacuum sputter chamber operating at $5x10^{-3}$ mbar Ar pressure at ambient temperature. A 15 nm $HfO_{2-x}$ metal oxide layer is formed using a 350 mJ and 10 Hz pulse KrF (248 nm) laser (Compex 205) Pulse Laser Deposition (PLD) system at ambient temperature following a comparable second layer photolithography technique. The top electrode (TE) underwent final photolithography, and 15 nm Mo was deposited in the sputter. Lithography steps included photoresist removal with acetone and cleaning techniques. The devices are inspected using a Hitachi TM4000Plus scanning electron microscopy (SEM) system at various fabrication phases. The SEM and picture of the last device used in this work are shown in Figures 1b and 1c, emphasizing the areas that overlapped for the electrical measurements. The finished thin film stack and the device geometry are shown in Figure 1a.

In electrical characterizations, a Keithley 2450 source meter and an AFG 31102 model Arbitrary Function Generator (AFG) were used to measure the DC transport Figure 3, LTP, LTD, and PPF of the memristor. The Mo top electrode was connected to the positive output (and negative output)to be applied 5.5V (and -5.5 V) pulses respectively. Instantaneous passing current through the memristor is read by the Keithley 2450 source-meter.

## RESULTS AND DISCUSSION

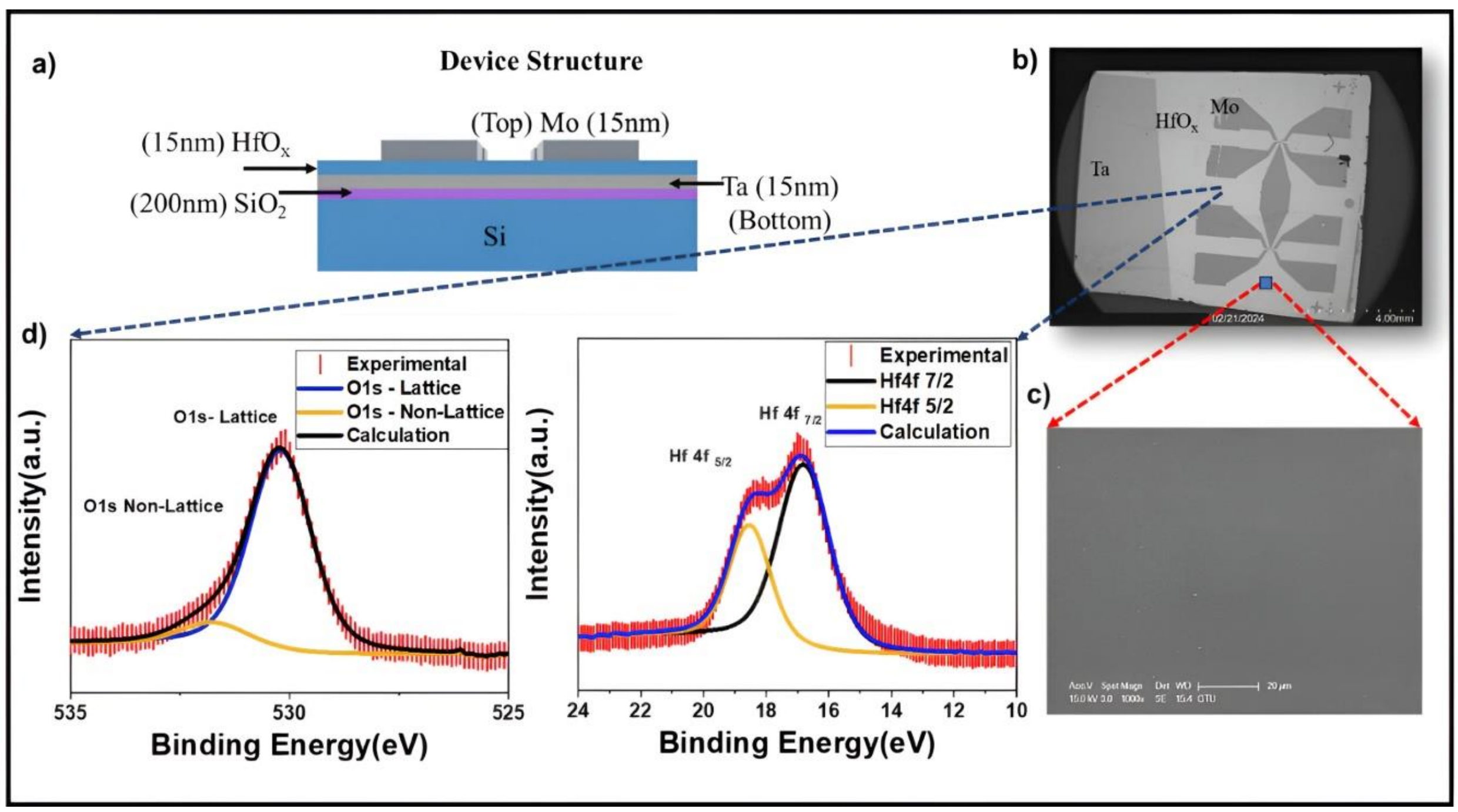

**Figure 1: a)** Device structure**, b-c)** SEM image of device (label 4 mm) and and $HfO_{2-x}$ surface (label 20 µm), **d)** The sample's XPS spectra for O1s (Left) and Hf4f (Right) are shown.

Figure 1a illustrates the layer structure of the memristor, where the top electrode (Mo) serves as a charge injector, enabling resistive switching dynamics of the device. The $HfO_{2-x}$ switching layer (15 nm) is the core functional region, where oxygen vacancies migrate to enable the SET and RESET states. The bottom electrode (Ta, 15 nm) acts as an oxygen vacancy reservoir, stabilizing the device's switching behavior. These layers are deposited on a $SiO_2$ substrate, which provides structural support and electrical isolation. High-resolution (Phoibos 100, SPECS GmbH, Berlin, Germany) spectroscopy has been utilized to examine the stoichiometry and degree of oxidation of the deposited layers on the sample surface. Figure 1d provides a detailed investigation of the Hf 4f and O1s peaks. Based on the XPS spectrum, the observed Hf 4f doublet peaks at 16.9 eV (4f 7/2) and 18.5 eV (4f 5/2)
indicate the presence of Hf 4f in $HfO_{2-x}$. In equivalent XPS of $HfO_2$ data, a spin-orbit coupling of 1.6 eV was found. [35-36]. These peaks were identified as the lattice and non-lattice oxygen, with binding energies of 530.3 and 531.8 eV, respectively, based on the best fitting of the O1s spectra. The quantitative analysis yields around 42% Hf and 58% O for $HfO_{2-x}$, indicating the growth oxide's stoichiometry.

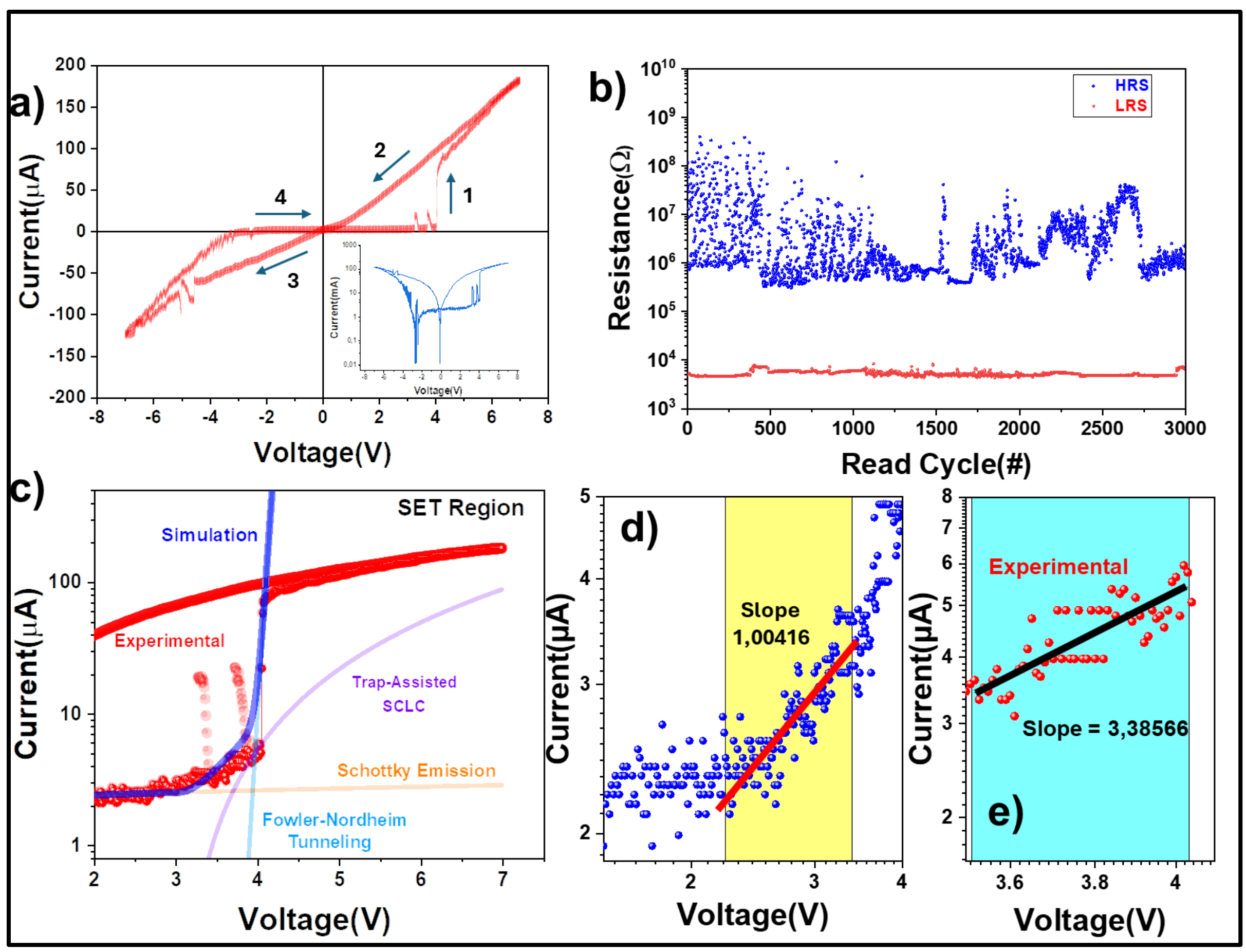

**Figure 2: a)** Current – voltage curve of the device, **b)** Retention test for HRS and LRS with 3000 reads (1 $V_{read}$), **c)** Set-state fitting of the device using Fowler-Nordheim (F-N) tunneling, space-charge-limited conduction (SCLC), and Schottky emission models, **d-e)** Linear fit of the current – voltage curve from 2.2 V to 3.4 V and, from 3.5 V to 4.03 V, indicating ohmic and trap-assisted SCLC behavior with the slope value included.

Figure 2a, the pinched hysteresis loop is the characteristic signature of memristor behavior [15]. It shows SET and RESET transitions of the memristor in current-voltage (I-V) graphs with linear sweeping voltages. The scan sequence was 0 → + 7.5 V → 0 → - 7.5 V → 0. Direction (1) represents the positive voltage sweep during which the device undergoes the SET process, switching from the high resistance state (HRS) to the low resistance state (LRS). Direction (2) corresponds to the initial part of the downward sweep, where the device remains in LRS. Direction (3) shows the RESET process occurring under negative voltages, where the device switches back from LRS to HRS via analog switching mechanism. Direction (4) represents the return sweep toward positive voltages, during which the device remains in the HRS.

The resistive switching mechanism makes memristors particularly attractive. Since memristors can switch between an HRS and an LRS [39]. The LRS is formed when oxygen vacancies align to construct a conductive bridge between the Ta and Mo electrodes. The HRS corresponds to the rupture of this conductive bridge, where the oxygen-vacancy filament ruptures. HRS is a stable, non-volatile state that does not require a continuous power supply, making it essential for data retention. HRS is the (RESET) off-state, 0 in the binary data storage system, while LRS is the (SET) on-state, 1 in the binary data storage system [38].

As shown in Figure 2b, the SET and RESET states were read 3000 times using a 1 V read voltage. LRS and HRS remained stable, and from this graph, the $R_{off}/R_{on}$ value was determined to be approximately $10^2$. The variation in the HRS is attributed to measurement noise inherent in high-resistance readings using a source meter. At high resistance levels, the measured current approaches the instrument's detection limit, making it more susceptible to fluctuations. The fact that variations in LRS and HRS follow correlated patterns further justifies the reliability of the measured data, indicating that the observed fluctuations stem from inherent device behavior rather than random noise.

Figure 2c illustrates the conduction mechanisms in the set state of the device, derived from the curve fitting of the first quadrant of the I - V hysteresis loop. The identified mechanisms include trap-assisted space-charge-limited conduction (SCLC – purple line), Fowler-Nordheim (F-N) tunnelling (cyan line), Schottky emission (orange line) mechanisms. A simulated switching curve (blue line) represents the cumulative effect of these mechanisms, modeled according to the Equation (1).

$$I_{total}(V) = I_{ohm}(V) + I_{schottky}(V) + I_{FN}(V) + I_{trap-assisted\ SCLC}(V)$$

$$= GV + AA^{*}T^{2}e^{-\frac{q(\emptyset_B - \Delta\emptyset)}{k_B T}} + aV^{2}e^{-\frac{b}{V}} + \frac{9}{8}\varepsilon_r\varepsilon_0\mu\frac{V^2}{d^3}\theta A \quad (1)$$

Schottky emission is the dominant current conduction mechanism observed until 2.1 V. The analysis in Figure 2c indicates that in this range, the current conduction occurs via the thermionic emission mechanism over the potential barrier between the Ta electrode and $HfO_{2-x}$. Beyond the Schottky emission continues to contribute to the current within a certain current range. In Figure 2d, the slope of the I-V curve between 2.2 V and 3.4 V which is 1.00416 indicates an ohmic current contribution in this region, suggesting that after Schottky emission, the ohmic mechanism takes over as the primary conduction process. In this voltage range, the linear I–V relationship confirms that the current is governed by Ohm's law. In Figure 2c, a noticeable transition occurs beyond 3.78 V. In the voltage range between 3.78 V and 4.03 V, is dominated by trap-assisted SCLC. This suggests that, at higher voltages, mechanisms such as trap-assisted SCLC start to govern the charge transport, replacing the previously dominant ohmic behavior. This indicates that beyond this range of voltage, carrier injection becomes efficient enough to fill available trap states, leading to a bulk-limited transport regime governed by space charge effects within the $HfO_{2-x}$ layer. It is necessary to determine whether the SCLC current behavior is trap-free or trap-assisted. For this purpose, in Figure 2e, SCLC behavior was graphically analyzed in the voltage range of 3.5 – 4.03 V. A linear fit was applied to the logarithmic plot of the data obtained within this voltage range. The fact that the slope value in the 3.5 – 4.03 V range in Figure 2e is 3.38566 shows that it is above this threshold and that the SCLC behavior is trap-assisted. This indicates that, instead of moving freely within the material, the carriers are first captured by energy traps and then released, meaning the conduction is limited by the trap-filling process. This analysis shows that the conduction characteristic of the memristor is shaped not only by classical SCLC but also by trap dynamics. In Figure 2c, a sharp increase in current in the voltage range of 3.91 – 4.09 V is observed. This was fitted via Fowler-Nordheim (F-N) tunneling mechanism. In this voltage range where F–N tunneling occurs, the system switches to the LRS, marking the formation of a stable conductive path across the $HfO_x$ layer. Starting from 3.98 V, the F–N tunneling mechanism, which is the main current contributor, increases its current contribution to $72.81\mu A$ at 4.09 V, which corresponds to approximately 89% of the total current in the system at 4.09 V.

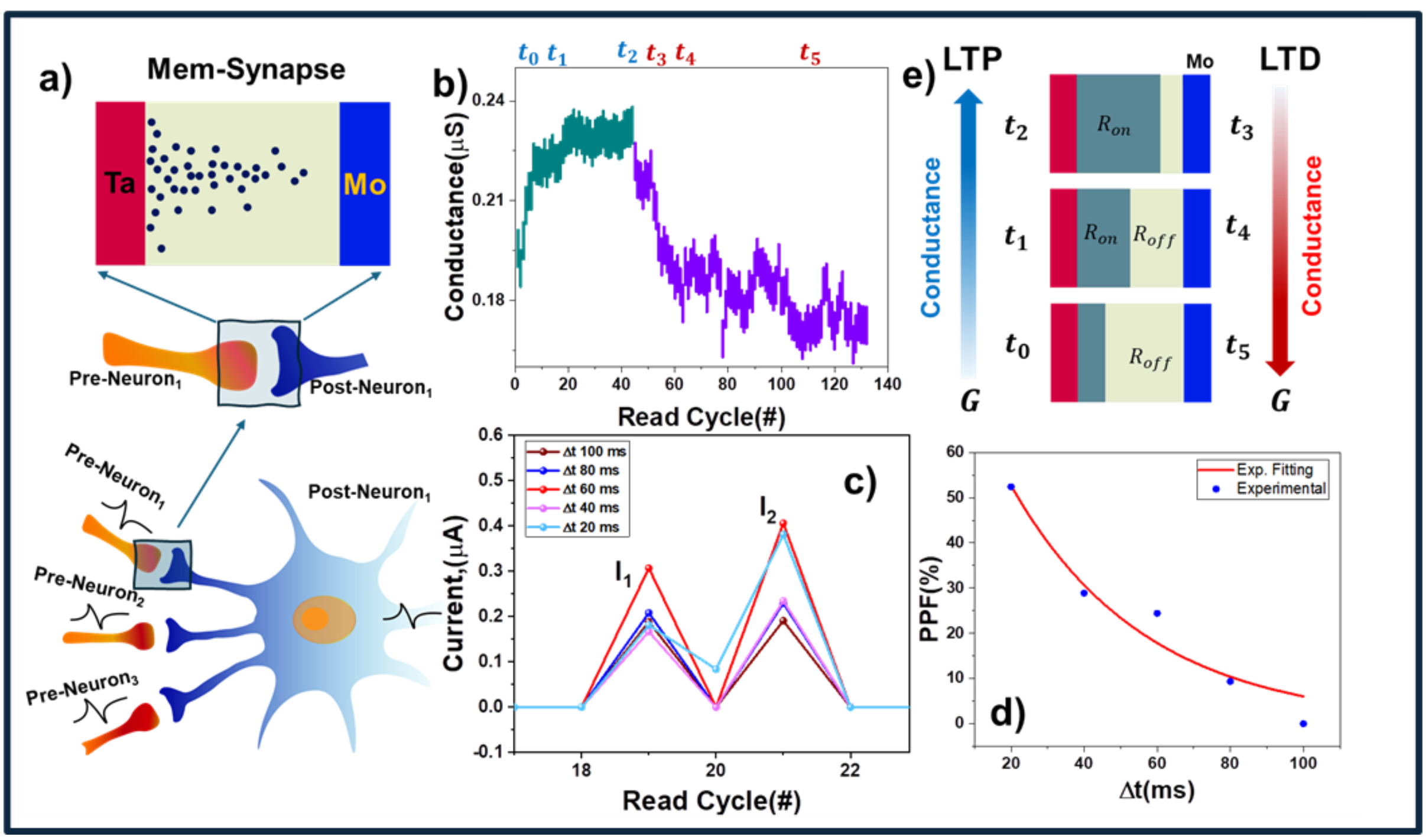

**Figure 3: a**) Illustration of the memristive synapse and interconnected pre-/post-neurons, showing the analogy between neural and memristive systems. **b)** During analog switching, the device conductance gradually increases with 44 potentiation pulses and decreases with 88 depression pulses (absolute value of depression). **c)** Paired-pulse facilitation (PPF) characteristics of the device with pulse interval (Δt) varying from 20 ms to 100 ms. **d)** Exponential fitting of experimental PPF data extracted from (c). **e)** Illustration of long-term potentiation (LTP) and depression (LTD) processes showing conductance modulation between $R_{off}$ and $R_{on}$ according to the time points given in (b).

Synaptic plasticity is the capability of changing the strength of synapses, depending on neural activity. This process allows for learning, memory, and adaptation skills in the human brain [21]. The synaptic weight defines the strength between synapses. If the synaptic weight increases, it means the long-term potentiation (LTP) phase, and it refers to learning and memory formation. On the other hand, a decrease in synaptic weight means long-term depression (LTD), and it refers to weakening the memory formation, forgetting [41]. Firstly, 44 pulses were applied with amplitude of 5.5 V. The pulse period was 300 ms, and the pulse width was 290 ms. Then, 88 more pulses were applied with an amplitude of -5.5 V in the same pulse period and width. These steps show how the conductance shifts between its maximum and minimum values under the 5.5 V pulses, as observed on a pulse-by-pulse basis. That demonstrates the synaptic reactions of the memristor to the given electric stimulus. Positive pulses led the device to the LTP (learning phase) while negative pulses led to the LTD (forgetting phase) [41].

Paired-Pulse Facilitation (PPF) is a process that simulates the time course of synaptic efficacy, and memristors can mimic this synaptic behavior [42]. The change in PPF percentage is representative of natural synaptic functioning and shows the temporal trend of the signal occurring across synapses. In that way, 2 pulses were applied with an amplitude of 5.5 V to the memristor with a constant pulse period of 300 ms. The pulse width was decreased by 20 ms in five successive steps. Increasing the difference between the pulse period and width means the time difference between two consecutive stimuli. If "$\Delta t$" increases, the intensity of the second pulse will be less. After the first pulse is applied, the carrier charges between the electrodes in the memristor and the ions remain active. However, these carrier charges and the effects of the ions on the memristor become passive over time after the applied stimulus. At this point, the shorter the time between the first and second pulses, the more the carrier charges and ions create stimulus intensity on top of the effect that is produced by the active carrier. [43] This also shows that if $\Delta t$ decreases, the stimulus intensity increases. Experimentally, this behavior is visualized as sequential current pulses, as seen in Figure 3c. The experimental output is plotted as PPF percentage in Figure 3d. At this point, the PPF percentage decreases exponentially compared to "$\Delta t$".

$$PPF = \left(\frac{I_2 - I_1}{I_1}\right) \times 100\% = ae^{-\frac{\Delta t}{\tau}} \qquad (2)$$

$$a = 90.53 \qquad \tau = 36.955\ ms$$

In Equation (2), $I_1$ and $I_2$ represent the current responses generated by the first and second pulses, respectively. It denotes the exponential fitting parameter of PPF percentage graph with respect to experimental data. "$a$" is fitting coefficient while "$\tau$" is response time of memristor conductance

Although no degradation was observed in terms of memristive behavior during electrical characterization measurements, the data obtained from successive measurements revealed small but significant changes in the behavior of the memristive element over time, in the sequence continuing from the I-V measurement to the synaptic and the PPF measurements. A gradual decrease observed in the current values indicated that the internal dynamics of the memristor changed over time and led to the interpretation of a gradual increase in the resistance value of the device. This increase in resistance can be attributed to changes occurring over time in the physical or chemical properties of the memristive structure as a result of high voltage and highly repetitive measurements. The observation of memristive characteristics in each measurement and the fact that the current value only decreased over time shows that this situation is a natural consequence of the device, and the fact that it did not hinder the tendency of the device's memristive behavior can also be considered as a supporting point. This situation should be evaluated by taking into account the dynamic nature of the device along with its application-dependent performance variability.

**Table 1:** Comparison between our device and the devices reported in the literature.

| Device | Metal- Oxide Thickness (nm) | $V_{set}(V_{reset})$ $[V]$ | $R_{off}/R_{on}$ | Retention $[s]$ | References |
|---|---|---|---|---|---|
| $Ta/\ TaO_x/\ HfO_2\ /\ Pt$ | 18.1 | 4(-2) | ~$10^2$ | >$10^4$ | [54] |
| $Ta/\ HfO_2\ /\ Pt$ | 5 | 2.2(-4) | N/A | $2.7 \times 10^6$ | [35] |
| $Pt/\ HfO_{2-x}\ /\ Ti$ | 20 | 4(-5) | ~10 | $10^4$ | [55] |
| $Ag/\ HfO_2\ /\ Pd$ | 6 | 1.1(-0.5) | ~$10^8$ | $10^5$ | [56] |
| $Mo/\ SiO_2\ /\ W$ | 19 | Analog (Analog) | N/A | >$10^5$ | [48] |
| $W/WO_{3-x}/HfO_2/\ Pd$ | 23 | 3.3(-3.5) | ~10 | $10^4$ | [53] |
| $Pt/HfO_2/Al_2O_3\ /TaN$ | 5 | -2.7(1.9) | ~$10^6$ | $10^4$ | [50] |
| $Pt/HfO_2/SiO_2\ /TaN$ | 7 | -2.6 ~ -1.8 (1.9)~(2.6) | ~$10^4$ | $10^4$ | [51] |
| $TiN/HfO_2/HfO_2{:}Al/Pt$ | ~8 | N/A | ~$10^2$ | $10^4$ | [52] |
| $Ta/HfO_2\ /Mo$ | 15 | 4.04 (Analog) | ~$10^2$ | $3 \times 10^3$ | This work |

Table 1 presents a comparison of the devices reported in the literature and this work according to the corresponding parameters. The comparison is based on parameters which are the metal-oxide thickness, set and reset voltage, Roff / Ron ratio, and the retention time. Roff / Ron ratios are based on the calculations using the high-resistance state (HRS) and low-resistance state (LRS) values obtained from retention tests. For devices that do not exhibit a distinct SET or RESET voltage, those showing analog switching behavior, the notation “(Analog)” is used in the relevant column. For devices where set and reset voltage values vary depending on the number of measurements, the format “Vset ~ Vset (Vreset) ~ (Vreset) is adopted; for approximate values, the “~” symbol is placed before the numbers. Values reported only as greater or smaller in the literature are denoted using the mathematical symbols “>” and “<”, while unavailable data are indicated as “N/A”.

## CONCLUSION

In this study, a memristor device in Ta/$HfO_{2-x}$/Mo structure was fabricated using magnetron sputtering and pulsed laser deposition, and supported by SEM imaging. The oxygen defects were determined by the Hf/O ratio using the XPS method. The XPS results confirmed the suitability of the device for resistive memory applications. Memristive characteristics and retention values were determined through DC transport measurements. The device exhibited stable resistive switching and clear HRS/LRS separation. A Python-based program was used to identify the contributions of Schottky emission, ohmic conduction, trap-assisted SCLC, and F–N tunneling to the total current. LTP-LTD and PPF measurements were performed to evaluate synaptic behavior. The effects of the two oxidizable materials used in the memristor device were successfully demonstrated in terms of both synaptic performance and resistance state stability.

**Acknowledgment**

This work was supported by the Scientific and Technical Research Council of Turkey (TUBITAK) through the project project No. 121F390.

**Author contributions:**

SK, BÖ and KK conceived of the presented idea. SK, BÖ and KK verified the analytical methods. The device has been fabricated by BÖ and methods. Device BÖ carried out the X-Ray photoelectron spectroscopy studies of the samples. AHC carried out device image by SEM. KK has all carried out electrical and synaptic measurements. Everyone who wrote a piece of the final manuscript discussed the outcomes.

**Conflicts of interest:** All authors declare that they have no conflicts of interest.

**Data and code availability:** All data produced by authors are presented in this manuscript and they are available in DOİ:

**Supplementary information:** There is no additional material omitted from the main body of the text.

**Ethical approval:** Ethical approval is not applicable, because this article does not contain any studies with human or animal subjects.

**ORCID IDs:**

Kerem Karataş :https://orcid.org/0009-0002-5834-2428

Bünyamin Özkal :https://orcid.org/0000-0002-9964-9250

Abdullah H. Coşar :https://orcid.org/0009-0000-5273-5736

Sinan Kazan :https://orcid.org/0000-0002-8183-5733